\documentclass[aip,showpacs,preprint,floatfix,preprintnumbers,amsmath,amssymb,superscriptaddress]{revtex4-1}

\usepackage{graphicx,color}
\usepackage{dcolumn}
\usepackage{amsmath}
\usepackage{amssymb}
\usepackage{amsfonts}
\usepackage{upgreek}
\usepackage{bm}

\begin{document}
\newcolumntype{d}[1]{D{.}{.}{#1}}
\title{Parity violation in nuclear magnetic resonance frequencies 
of chiral tetrahedral tungsten complexes NWXYZ (X, Y, Z = H, F, Cl, Br or I)}

\author{Sophie Nahrwold}
\email{nahrwold@fias.uni-frankfurt.de; present address:
Dotterbeek 7, NL-5501 BG Veldhoven, The Netherlands}%
\affiliation{%
Frankfurt Institute for Advanced Studies, Goethe-University Frankfurt am Main, Ruth-Moufang-Str. 1,
D-60438 Frankfurt am Main, Germany}%
\affiliation{%
Clemens-Sch\"opf-Institute, Technical University Darmstadt, Petersenstr. 22, D-64287 Darmstadt, Germany}%

\author{Robert Berger}
\email{r.berger@fias.uni-frankfurt.de}
\affiliation{%
Frankfurt Institute for Advanced Studies, Goethe-University Frankfurt am Main, Ruth-Moufang-Str. 1,
D-60438 Frankfurt am Main, Germany}%
\affiliation{%
Clemens-Sch\"opf-Institute, Technical University Darmstadt, Petersenstr. 22, D-64287 Darmstadt, Germany}%

\author{Peter Schwerdtfeger}
\email{p.a.schwerdtfeger@massey.ac.nz}
\affiliation{%
Centre for Theoretical Chemistry and Physics, The New Zealand 
Institute for Advanced Study, Massey University Albany, Private Bag 102904, North Shore City,
Auckland 0745, New Zealand}
\affiliation{%
Fachbereich Chemie, Philipps-Universit\"{a}t Marburg, 
Hans-Meerwein-Str., D-35032 Marburg, Germany}

\date{\today}

\begin{abstract}
Density functional theory within the two-component quasi-relativistic
zeroth-order regular approximation (ZORA) is used to predict parity
violation shifts in ${}^{183}$W nuclear magnetic resonance shielding
tensors of chiral, tetrahedrally bonded tungsten complexes of the form
NWXYZ (X, Y, Z = H, F, Cl, Br or I). The calculations reveal that sub-mHz
accuracy is required to detect such tiny effects in this class of compounds, 
and that parity violation effects are very sensitive to the choice of ligands.
\end{abstract}

\maketitle

\section{Introduction\label{sec:intro}}
Electro-weak currents violate parity symmetry, thus lifting the energetic
degeneracy of enantiomers which leads to tiny differences in molecular 
properties of non-identical mirror-image molecules.\cite{Yamagata66,Rein:1974,Gajzago:1974,Letokhov:1975,Kompanets} 
Proposed experimental schemes to detect PV shifts in properties of chiral
molecules range from M\"ossbauer spectroscopy,\cite{Compton02} to vibrational 
spectroscopy,\cite{Kompanets,arimondo:1977,bauder:1997,Daussy:1999,Chardonnet99,Montigny10,Saue13} 
electron paramagnetic resonance (EPR) spectroscopy,\cite{harris:1980,khriplovich:1985} rotational
spectroscopy,\cite{bauder:1997} nuclear magnetic resonance (NMR) spectroscopy\cite{gorshkov:1982,barra:1986,Barra88,barra:1988,Barra87,barra:1996,robert:2001,Laubender03}
including nuclear spin-spin couplings,\cite{Manninen05,Ledbetter09}
quantum beats in optical rotation,\cite{harris:1978} and finally tunneling 
dynamics of chiral molecules.\cite{Quack86,berger:2001a,berger:2003,Quack05,Gottselig01a}
Despite some great efforts to measure such tiny PV effects in chiral
molecules, which are expected to lie in the $\upmu$Hz to Hz region, 
all experiments have to be considered unsuccessful or unconvincing so far,
despite occasional claims to the contrary (for a details discussion see Ref.~\cite{Schwerdtfeger:10}).
The current status on experimental and theoretical attempts to search for
suitable chiral molecules and subsequent PV measurements has been reviewed
several times.\cite{Schwerdtfeger:10,Berger04,Crassous:2005,Quack08,Darquie10}

Experimental detection of nuclear magnetic resonance (NMR) frequency
shifts is considered to be a feasible route towards a first
observation of PV in chiral
molecules.\cite{gorshkov:1982,barra:1986,barra:1988,barra:1996,robert:2001} 
In contrast to nuclear spin-independent molecular PV effects aimed for
in high-resolution vibrational spectroscopy, which may face
difficulties in achieving an accuracy superior to high-precision
atomic experiments on heavy elements such as cesium,\cite{Wieman97,Wieman99,ginges:2004,Derevianko07} 
a successful measurement of nuclear spin-dependent PV effects promises to give
insight into hadronic weak interactions inside the nucleus.\cite{haxton}
The reason for this is that for most nuclei the
dominant contribution to the nuclear spin-dependent weak interaction
between electrons and nuclei\cite{gorshkov:1982,gorshkov:1982a} comes
from the nuclear anapole moment,\cite{zeldovich:1957,zeldovich:1958,zeldovich:1959} 
which is caused by parity violating interactions within the nucleus and which has only
been measured once before in $^{133}$Cs.\cite{wood:1997}

The special electronic situation in open-shell diatomic molecules\cite{labzovsky:1977,labzovsky:1977a,labzovsky:1978,labzovsky:1978a,sushkov:1978,kozlov:1995}
offers in principle unmasked access to nuclear spin-dependent PV
interactions in a limited set of nuclei and promising candidate
systems such as RaF\cite{isaev:2010,isaev:2012,isaev:2013} have been
proposed for this research line. Due to the near degeneracy of levels
opposite parity in conventionally chiral molecules, the leading PV
contribution (nuclear spin-independent or nuclear spin-independent)
can conveniently be selected by the type of spectroscopy (NMR, MW, IR,
UV). Furthermore, one can draw in principle from a wealth of NMR
active nuclei to form chiral molecules, for which PV effects would be
observable as parity violating frequency differences between
enantiomers.
\begin{center}
\begin{figure}[!htbp]
\begin{center}
\includegraphics[width=.3\linewidth]{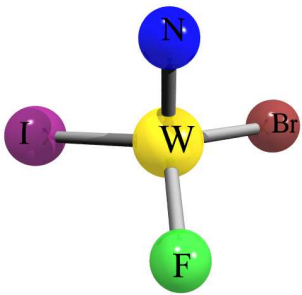}
\end{center}
\caption{(Color online) The S-enantiomer of the molecule NWBrFI.}
\label{fig:nwfbri}
\end{figure}
\end{center}
In this paper we compare PV effects in NMR shielding constants
obtained from two-component quasi-relativistic density functional
calculations for a whole range of chiral tetrahedral NWXYZ molecules
as shown in Figure \ref{fig:nwfbri} for the S-enantiomer NWBrFI, with X,
Y and Z representing hydrogen or the halogens F, Cl, Br and I. These
chiral molecules contain heavy elements and serve as model compounds
to estimate nuclear spin-dependent PV effects. However, we point out
that only the non-chiral species NWH$_3$ and NWF$_3$ have been
identified in the gas phase by Wang et al.\cite{wang:2008b,wang:2008a}
so far. Gas-phase syntheses of the
chiral derivatives with subsequent mass selection and isolation could
open up the way for high-resolution gas-phase NMR spectroscopy. We
mention that nuclear spin-independent PV effects in vibrational
spectra of these chiral tungsten derivatives have been studied
theoretically before.\cite{figgen:2010,Figgen:2010a}

\section{\label{sec:theory}Methodology}
The main contributions to the PV operator $\hat{H}_{\mathrm{PV}}$ 
for electron-nucleus interaction
consists of a nuclear spin independent $\hat{H}^{\left( 1\right) 
}_{\mathrm{PV}}$
and dependent $\hat{H}^{\left( 2\right) }_{\mathrm{PV}}$ part,
\begin{align}
\hat{H}_{\mathrm{PV}}&=\hat{H}^{\left( 1\right) 
}_{\mathrm{PV}}+\hat{H}^{\left( 2\right) }_{\mathrm{PV}} \nonumber \\
&= 
\frac{G_{\mathrm{F}}}{2\sqrt{2}}\sum_{i}\left(\sum_{A=1}^{N_{\mathrm{nuc}}} 
\, Q_{\mathrm{W}}\left( A\right) \gamma^5_i \varrho_A\left( 
\vec{r_i}\right)  +\sum_{A=1}^{N_{\mathrm{nuc}}} \kappa_{A} 
\varrho_A\left( \vec{r_i}\right) 
\vec{\alpha_i}\cdot\vec{I}_{A}\right) ,
\label{eq:diracpvop}
\end{align}
where $G_{\mathrm{F}}=(2.22254\times 10^{-14})E_{\mathrm{h}}a_{0}^{3}$
is Fermi's weak coupling constant, and the sum runs over all
$N_{\mathrm{nuc}}$ nuclei in the system. $Q_{\mathrm{W}}\left(
A\right)\approx\left(1-4 \sin^2\theta_\mathrm{W} \right)Z_A-N_A $ is
the electroweak charge of nucleus $A$ with proton number $Z_A$ and
neutron number $N_A$. In the present work, the value for the Weinberg
angle is set to $\sin^2\theta_\mathrm{W}=0.2319$ as used in previous
calculations. A recent value determined at energies comparable to
those considered here is
$\sin^2\theta_\mathrm{W}=0.2397\left(18\right)$.\cite{anthony:2005}
We mention that radiative corrections alter the weak charge and one
approximately obtains $Q_{\mathrm{W}} \approx -0.9857N_A + 0.0675Z_A$.\cite{Milstein02,Johnson07} 
$\rho_{A}(\vec{r})$ and $\vec{I}_A$ are the normalised nucleon density and the nuclear spin, respectively,
$\gamma^5$ is the Dirac pseudoscalar (chirality operator),
$\vec{\alpha_i}$ is a vector comprised of the Dirac matrices in
standard form, $\vec{r_i}$ is the electron position vector and
$\kappa_{A}$ is a nuclear state dependent parameter. We note here that
different choices for the constant related to the nuclear spin-dependent 
parameter can be found in the literature, for instance $\kappa_{A}$,\cite{nahrwold:2009}
$k_{\mathcal{A},A}$,\cite{isaev:2012}, $g_{2A}^P$,\cite{gorshkov:1982,kozlov:1995} 
$k_{\mathrm{NSD},A}$,\cite{Borschevsky:2012} 
and $(-\lambda)(1-4\sin^2\theta_\mathrm{W})$,\cite{barra:1986,barra:1988,Laubender03,Soncini03,weijo:2005,Laubender06}
which are simply related by
\begin{align}
\kappa _{A}/2 = k_{\mathcal{A},A} = g_{2A}^P = 2 k_{\mathrm{NSD},A} = 
-\left(1-4\sin^2\theta_\mathrm{W} \right) \lambda_A ,
\label{eq:anapole}
\end{align}
where $\lambda_A$ is a nuclear state (spin) dependent parameter. 
There are two contributions to these parameters.
The first contribution comes from the 
electroweak neutral coupling between electron vector and nucleon
axial-vector currents ($\vec{V}_e\vec{A}_N$),\cite{Novikov77}
and the second comes from the nuclear anapole moment. The nuclear anapole 
moment scales with $\kappa_A \sim A^{2/3}$ and becomes the dominant 
contribution for heavy nuclei (see Refs.~\onlinecite{Flambaum84,ginges:2004,dmitriev:2000,isaev:2010,Borschevsky:2012} 
for details).

The first part $\hat{H}^{\left( 1\right) }_{\mathrm{PV}}$ of the PV
operator comprises the dominant contribution from the first-order
neutral current interaction between electrons and nuclei. When nuclear
spin-independent molecular properties, such as vibrational frequencies
or electronic energies are considered, this is usually the only part
of the operator taken under consideration in 
calculations.\cite{Laerdahl99,berger:2000a,quack:2000,quack:2001,berger:2001,schwerdtfeger:2002,berger:2003,quack:2003a,stohner:2004,schwerdtfeger:2004b,fokin:2006,berger:2007}.
For explicitly nuclear spin-dependent properties such as NMR
frequencies or hyperfine splittings, however, the second part of the
operator, $\hat{H}^{\left( 2\right) }_{\mathrm{PV}}$, is estimated to
yield the dominant PV contribution in most cases.\cite{gorshkov:1982}
It should be kept in mind, however, that under certain conditions the
nuclear spin independent operator can contribute to the interaction
with similar strength.\cite{flambaum:1993} Furthermore, because of the approximate
$Z^2$-scaling of $\hat{H}^{\left( 2\right) }_{\mathrm{PV}}$ operator,
it is important to study nuclear spin-dependent PV effects in chiral
molecules containing heavy elements.\cite{Laubender03,Soncini03,bast:2006,Laubender06}

Through the factor $\lambda_A$, $\hat{H}^{\left( 2\right)}_{\mathrm{PV}}$ 
contains also higher-order nuclear weak interaction
effects. Because of the uncertainty for this factor (as it is of
nuclear structure origin) we set $\lambda_A = -1$ in all our
calculations and therefore values reported herein for NMR shielding
constants and frequency splittings are effective in the sense that
they have to be scaled by the negative of the actual value of
$\lambda_A$, in order to obtain an estimate of measurable physical
values. For heavy nuclei, it is expected to lie between 1 and
10.\cite{flambaum:1980,flambaum:1984}

All parity violation (PV) calculations were performed within the
framework of the two-component quasi-relativistic zeroth-order
regular approximation (ZORA),\cite{chang:1986,lenthe:1993,lenthe:1994,lenthe} which was applied by
Berger et al.\cite{berger:2005,berger:2005a} to the calculation of PV
energy differences between enantiomers and later\cite{nahrwold:2009}
to first-order PV contributions to NMR shielding tensors. Within this
approach, the  perturbed Hamiltonian including external magnetic field
effects and PV contributions up to first order is given by (for
simplicity we omit summation over electrons)
\begin{align}
\hat{h}^{\mathrm{zora}}
&= \vec{\sigma} \cdot \vec{p} \, \tilde{\omega} \,   \vec{\sigma} 
\cdot \vec{p} + V
-q \left\lbrace \vec{\sigma} \cdot \vec{p} , \tilde{\omega} 
\vec{\sigma} \cdot \vec{A}\right\rbrace  +  q^2 \left\lbrace 
\vec{\sigma} \cdot \vec{A}, \tilde{\omega} \vec{\sigma} \cdot \vec{A} 
\right\rbrace + \frac{1}{c} \left\lbrace \vec{\sigma} \cdot \vec{\pi} 
, \tilde{\omega} \hat{f}_{\mathrm{PV}} \right\rbrace  ,
\label{eq:zorahamiltonian}
\end{align}
with $\left\lbrace x , y \right\rbrace=xy+yx$ being the
anticommutator. Here, $\vec{p}$ is the linear momentum of the
electron, $\vec{\pi} = \vec{p} -q \vec{A}$ the conjugate momentum
operator inside a magnetic field with vector potential $\vec{A}$, and
$q=-e$ the charge of the electron. $c$ is the speed of light in vacuum
and $\vec{\sigma}$ is the vector containing the three Pauli matrices.
The nuclear spin-dependent PV operator $\hat{f}_{\mathrm{PV}}^{(2)}$ in the
two-component form enters the Dirac (2$\times$2) matrix equation in
the off-diagonal and can be written as,
\begin{align}
\hat{f}_{\mathrm{PV}}^{(2)} = 
\lambda_{\mathrm{PV}}\sum_{A=1}^{N_{\mathrm{nuc}}}\frac{\kappa_A}{\hbar 
\gamma_A} \varrho_A\left( \vec{r}\right) 
\vec{\sigma}\cdot\vec{\mu}_{A} ,
\label{eq:zorapvop}
\end{align}
where we introduced $\lambda_{\mathrm{PV}}$ as a formal perturbation 
parameter, $\hbar= h/2\pi$ is the reduced Planck constant, and 
$\gamma_A$ and $\vec{\mu}_{A}=\hbar  \gamma_A \, \vec{I}_A$ are the 
gyromagnetic ratio and magnetic moment of nucleus $A$ respectively.

Within the density-functional theory (DFT) approximation pursued 
here, the potential $V$ is given by $V = V_{\mathrm{H}} + 
V_{\mathrm{XC}} + V_{\mathrm{N}}$ with the Hartree potential 
$V_{\mathrm{H}}$, the exchange-correlation potential 
$V_{\mathrm{XC}}$ and the electron-nucleus attraction potential 
$V_{\mathrm{N}}$. The ZORA factor $\tilde{\omega}$ is given by,
\begin{equation}
\tilde{\omega} =\frac{1}{2m_{\mathrm{e}}-V/c^{2}},
\label{eq:omega}
\end{equation}
and computed using van W{\"u}llen's model potential $\widetilde{V}$
with additional damping instead of the 
actual potential $V$ in order to circumvent a direct dependence of 
$\tilde{\omega}$ on the electron orbitals and to alleviate the gauge 
dependence of the ZORA approach.\cite{wullen:1998,liu:2001}  
The model potential $\widetilde{V}$ used in the calculation of the ZORA factor $\tilde{\omega}$ of Eq.~\ref{eq:omega} is described in detail in Ref. \onlinecite{wullen:1998}. Model potentials are calculated using the local density approximation exchange--correlation functional and superpositions of atomic model densities $\tilde{\rho}^{\mathrm{mod}}_A$. These model densities are expanded in terms of Gaussian functions as described in Ref. \onlinecite{wullen:1998}. Exponents  $\alpha_{iA}^{\mathrm{mod}}$ and coefficients $c_{iA}^{\mathrm{mod}}$ used in this work were determined by Christoph van W\"ullen \cite{wullen:1998} and are listed in Table~S1.

In ZORA SCF calculations and for the parity violating interaction, we used a Gaussian nuclear model\cite{visscher:1997} with exponent coefficients
\begin{equation}
\alpha_{\mathrm{nuc}}= \frac{3}{2 r_{\mathrm{nuc}}\left(A\right)^{2}}
\end{equation}
and 
\begin{equation}
r_{\mathrm{nuc}}\left( A\right) = \left(0.836 A^{1/3} + 0.570\right)\mathrm{fm},
\end{equation}
$A$ being the atomic mass number of the isotope with charge $Z$. For terms arising due to the vector potential $\vec{A}_{\mu}$, however, a point-like distribution of the nucleus' magnetic moment has been assumed. All calculations reported herein were performed with a modified version of the TURBOMOLE program.\cite{ahlrichs:1989,haser:1989}

The parity-conserving NMR shielding tensor for a given nucleus $Q$ is 
defined as the second derivative of the total energy $E$ with respect 
to the nuclear magnetic moment $\vec{\mu_{Q}}$ of the nucleus under 
study and the externally applied magnetic field $\vec{B}$ taken for 
vanishing perturbations,
\begin{align}
\sigma_{kt}\left(Q \right) :=\left. 
\frac{d^{2}E}{dB_{k}d\mu_{Qt}}\right|_{ \vec{T}=\vec{0}},
\end{align}
where $\vec{T}$ is the vector containing all perturbation parameters: 
$\vec{T}^{\mathrm{T}} = \left(\vec{B}^{\mathrm{T}}, 
\vec{\mu}_1^{\mathrm{T}}, \dots, 
\vec{\mu}_{N_{\mathrm{nuc}}}^{\mathrm{T}}, 
\lambda_{\mathrm{PV}}\right)$. Accordingly, the parity non-conserving 
NMR shielding tensor can be defined as a third-order derivative of 
$E$ with respect to $\vec{\mu_{Q}}$, $\vec{B}$ and the PV 
perturbation parameter $\lambda_{\mathrm{PV}}$,
\begin{align}
\sigma_{kt}^{\mathrm{PV}}\left(Q \right) :=\left. 
\frac{d^{3}E}{dB_{k}d\mu_{Qt}d\lambda_{\mathrm{PV}}}\right|_{ 
\vec{T}=\vec{0}}.
\label{eq:sigpv}
\end{align}

After taking derivatives and neglecting contributions due to the
nuclear spin-independent PV contribution $\hat{f}_\mathrm{PV}^{(!)}$, 
one arrives at the following expression for the PV NMR shielding 
tensor,\cite{nahrwold:2009}
\begin{eqnarray}
\sigma_{kt}^{\mathrm{PV}}\left(Q \right)&=& 
\sigma_{\mathrm{d},kt}^{\mathrm{PV}}\left(Q 
\right)+\sigma_{\mathrm{p},kt}^{\mathrm{PV}}\left(Q 
\right)+\sigma_{\mathrm{so},kt}^{\mathrm{PV}}\left(Q \right) 
\label{eq:sigzora2}\, .
\end{eqnarray}
Subscripts d, p, so refer to diamagnetic, paramagnetic and spin-orbit 
coupling contributions according to their conventional meaning,\cite{wolff:1999}
except for $\sigma_{\mathrm{p}}^{\mathrm{PV}}$ and 
$\sigma_{\mathrm{so}}^{\mathrm{PV}}$, where contributions were 
grouped together according to their scaling behavior with respect to 
the nuclear charge (see Ref. \onlinecite{nahrwold:2009} for details and
explicit expressions). 

Results will be reported in terms of NMR frequency splittings 
$\Delta\nu^{\mathrm{PV}}$ between left($S$)- and right($R$)-handed 
enantiomers inside a static homogeneous magnetic field of flux density $B$:
\begin{align}
\label{eq:freq}
\Delta\nu^{\mathrm{PV}}\left(Q \right) = \nu^S\left(Q \right) - 
\nu^R\left(Q \right) = B \gamma_{Q} \sigma^{\mathrm{PV}}\left(Q 
\right)/\pi ,
\end{align}
with the isotropic shielding constant $\sigma^{\mathrm{PV}}= 
\frac{1}{3}\mathrm{Tr}\left[\sigma^{\mathrm{PV}} \right]$, to which 
the traceless diamagnetic PV shielding tensor 
$\sigma_{\mathrm{d},kt}^{\mathrm{PV}}\left(Q \right)$ 
does not contribute. For certain choices of density 
functionals, the remaining two linear response type terms can be computed 
within an uncoupled Kohn--Sham framework as the corresponding response 
equations can be decoupled due to time-reversal symmetry.

\section{\label{sec:compdet}Computational Details}

Gaussian basis sets used in calculating the PV NMR shielding tensors for
the NWXYZ series of molecules discussed herein were all uncontracted and
were constructed as described in the following. For hydrogen, the
aug-cc-pVDZ  basis set of Ref.  \onlinecite{dunning:1989,kendall:1992} was
employed (in uncontracted form). Exponent coefficients
$\alpha_i^\mathrm{orb}$ of the uncontracted spherical Gaussian basis sets
for all other atoms were taken from an even tempered list generated
according to $\alpha_i^\mathrm{orb} = \gamma \beta^{N-i}$ with $\quad i =
1,2,\dots,N$ and $N=26$.  Herein we chose the largest exponent coefficient
as $\alpha_1^\mathrm{orb} = 500000000~a_0^{-2}$ and the smallest as
$\alpha_{26}^\mathrm{orb} = \gamma = (2/100)~a_0^{-2}$.
The set of exponents taken from this list are 
(1-25s,2-26p,20-24d,22-23f) for N, (1-25s,2-26p,15-25d,22-23f) for F, 
(1-25s,2-26p,15-25d,22-23f) for Cl,
(1-25s,2-26p,15-25d,20-24f) for Br, (1-25s,2-26p,12-25d,15-25f) for 
W, and (1-25s,2-26p,15-25d,20-24f) for I.

For the PV calculations the gradient corrected Becke-Lee-Yang-Parr
functional (B-LYP) was used.\cite{Becke1999,Lee1988} The structural
parameters for the compounds studied were obtained through
energy minimization using the B3LYP hybrid functional together with aug-cc-pVTZ basis sets, in connection with 
scalar relativistic pseudopotentials for W, Br and I, described in detail in
Refs.~\onlinecite{figgen:2010,Figgen:2010a}. Parity violating NMR shielding tensors were obtained 
for the $^{183}$W isotope (natural abundance 14.31\%), which has nuclear spin $I= 1/2$ and a
gyromagnetic ratio of 1.1282406$\times$10$^7$ rad T$^{-1}$ s$^{-1}$.\cite{handbook}
A common gauge origin placed at the position of the
tungsten nucleus was employed in this study. We note that extension of the 
basis set used leads to minor changes.  For example, extending the f-set for 
N to (20-24f) and for Cl to (20-24f) changes $\Delta\nu_{\mathrm{PV}}$ for 
$^{183}$W by only $-0.02$ $\upmu\mathrm{Hz}$ for NWBrClF compared to the smaller 
basis set. This is negligible compared to the effect different functionals 
would have on PV frequency shifts. For example, using the local density approximation\cite{LDA} 
we obtain (B-LYP values are given in parentheses) $-419$
($-363$) $\upmu\mathrm{Hz}$ for NWHFI, $-106$ ($-88$) $\upmu\mathrm{Hz}$ for
NWHClF, and $-106$ ($-94$) $\upmu\mathrm{Hz}$ for NWHBrCl for the $^{183}$W
NMR PV shift. The dependence of predictions for parity violation effects on 
the chosen density functional in vibrational spectra of these compounds has 
been discussed in detail before.\cite{Figgen:2010a}

\section{\label{sec:res}Results and discussion}

In order to facilitate the selection of a chiral compound especially well
suited for the investigation of PV NMR effects, it is important to assess
the effect of different nuclei surrounding the nucleus under study. PV NMR
frequency splittings $\Delta\nu_{\mathrm{PV}}$ for the $^{183}$W isotope
and energy differences $\Delta E_{\mathrm{PV}}$ between enantiomers for a
series of chiral molecules of the general structure NWXYZ with X,Y,Z =
H,F,Cl,Br or I have therefore been investigated and the results are listed
in Table \ref{tab:nwxyz}. The compounds studied herein are derived by
substitution from the NWH$_3$ and NWF$_3$ molecules synthesized by
Wang et al.\cite{wang:2008b,wang:2008a}
In the group of molecules from \textbf{1} to \textbf{4}
shown in Table \ref{tab:nwxyz} all three hydrogen atoms have been
substituted by halogens with respect to NWH$_3$, in the group from
\textbf{5} to \textbf{10} only two hydrogen atoms have been substituted.
Figure \ref{fig:nwfbri} shows, for example,  the structure of the
$S$-enantiomer of NWBrFI.

\begin{widetext}
\begin{center}
\begin{table}[h]
\begin{ruledtabular}
\newcolumntype{f}{D{.}{.}{0}}
\newcolumntype{e}{D{.}{.}{1}}
\caption{PV $^{183}$W NMR frequency splittings $\Delta\nu = \nu^S - 
\nu^R$ due to the isotropic parity violating NMR shielding constants 
and PV
energy differences $\Delta E_{\mathrm{PV}}$ between the $S$- and 
$R$-enantiomers
within a series of compounds NWXYZ with XYZ= H, F, Cl, Br or I, 
calculated using the B-LYP functional. The NMR frequencies
were obtained at a magnetic flux density of $B=11.7$~T and are given 
here in $\upmu$Hz. $\Delta E_{\mathrm{PV}}/h$ is given in Hz. The total 
PV frequency splitting $\Delta\nu_{\mathrm{PV}}$
is defined in Eq.~\ref{eq:sigzora2}. 
$\Delta\nu_{\mathrm{PV},\mathrm{p}}$ and 
$\Delta\nu_{\mathrm{PV},\mathrm{so}}$ are related to the 
paramagnetic
and spin-orbit coupling contributions to the isotropic part of the 
NMR shielding tensor.
Results are given with three significant figures for 
$\Delta\nu_{\mathrm{PV}}$, the individual contributions to 
$\Delta\nu_{\mathrm{PV}}$ were rounded
to the same accuracy as $\Delta\nu_{\mathrm{PV}}$.
\label{tab:nwxyz}}
\begin{tabular}{ll|d{4}d{4}d{4}d{4}}
Number & Molecule & \multicolumn{1}{c}{$\Delta\nu_{\mathrm{PV}}/\mathrm{{\upmu}Hz}$} & 
\multicolumn{1}{c}{$\Delta\nu_{\mathrm{PV},\mathrm{p}}/\mathrm{{\upmu}Hz}$}
& \multicolumn{1}{c}{$\Delta\nu_{\mathrm{PV},\mathrm{so}}/\mathrm{{\upmu}Hz}$} & 
\multicolumn{1}{c}{$\Delta E_{\mathrm{PV}}/(h~\mathrm{Hz})$}\\
\hline
\textbf{1} & NWBrClF & -9.09 & 1.15 & -10.24 & -27.9\\
\textbf{2} & NWClFI & -25.9 & 3.6 & -29.6 & -74.0\\
\textbf{3} & NWBrFI & -16.7 & 4.8 & -21.4 & -47.6\\
\textbf{4} & NWBrClI &  0.398 & 2.578 & -2.179 & -2.21\\
\hline
\textbf{5} & NWHBrCl & -94.1 & -43.4 & -50.8 & 83.4\\
\textbf{6} & NWHBrI &-196 &  -96 & -101 & 134\\
\textbf{7} & NWHClI & -293 & -139 & -154 & 214 ^{\left. a \right)}\\
\hline
\textbf{8} & NWHClF & -88.4 & -41.8 & -46.6 & -8.24 \\
\textbf{9} & NWHBrF & -176 &  -74 & -102 & 36.6\\
\textbf{10} & NWHFI & -363 & -152 & -210 & 104
\end{tabular}
\end{ruledtabular}
\begin{flushleft}
{$^{\left. a \right)}$ This value of $\Delta E_{\mathrm{PV}}/h$ compares to 
138 Hz (242 Hz) of Ref. \onlinecite{Figgen:2010a} obtained at the B3LYP (LDA) level of theory.}
\end{flushleft}
\end{table}
\end{center}
\end{widetext}

\begin{center}
\begin{figure}[!htbp]
\begin{center}
\includegraphics[width=.7\linewidth]{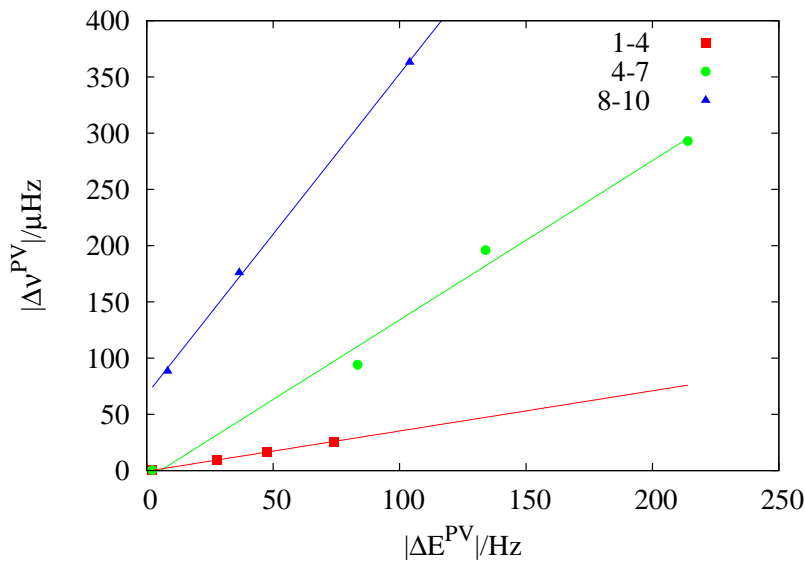}
\end{center}
\caption{(Color online) Plot of the PV $^{183}$W NMR frequency splitting 
$\Delta\nu_{\mathrm{PV}}$ against the PV energy difference $\Delta 
E_{\mathrm{PV}}$ (absolute values only)}
\label{fig:NMR}
\end{figure}
\end{center}

Beside the scaling of PV effects with nuclear charge, there is still 
no generally applicable simple model for PV effects available to design good
ligands around a central chiral atom for future PV measurements,\cite{Saue11}
and one has to rely on explicit quantum theoretical calculations.
However, Figure \ref{fig:NMR} indicates that there is a correlation 
between the absolute values of $\Delta\nu_{\mathrm{PV}}$
and $\Delta E_{\mathrm{PV}}$, and they group together in certain sets 
of molecules, i.e. those which contain no hydrogen, those
which contain fluorine and hydrogen and the remaining three 
containing hydrogen but no fluorine.
We observe the strongest PV energy difference for NWHClI, but we 
mention here that the results could be quite
sensitive to the DFT approximation applied.\cite{Visscher05}
Nevertheless, the order of magnitude of the PV energy differences $\Delta E_{\mathrm{PV}}$ is 
comparable to those reported for PbHBrClF,\cite{laerdahl:2000a}
and as expected much larger compared to the chiral polyhalomethanes
(CHXYZ)\cite{laerdahl:2000a,berger:2007} or polyhalocubanes,\cite{fokin:2006}
if one omits the hypothetical astatine derivative CHAtFI.

In the first series of compounds containing no hydrogen and showing rather
small PV effects, the relative ordering of the absolute values of the PV
frequency splittings $\Delta\nu_{\mathrm{PV}}$ and energy differences
$\Delta E_{\mathrm{PV}}$ is the same, i.e. 
$\mathbf{4<1<3<2}$.  Here, the most significant effect seems to be an order
of magnitude increase in $\Delta E_{\mathrm{PV}}$ and an even stronger
increase in $\Delta\nu_{\mathrm{PV}}$ upon introducing fluorine as a
ligand. The large impact of fluorine substitution is most probably due to
its large electronegativity which seems to cause a stronger ``chiral
field'' around the tungsten atom.  For PV energy shifts this has been
observed before.\cite{gottselig:2005} The paramagnetic and spin-orbit
coupling contributions to the frequency splitting,
are of opposite sign, with the negative spin-orbit coupling contribution
being larger by an order of magnitude with respect to the paramagnetic one
in compounds \textbf{1}, \textbf{3} and \textbf{2} and of roughly the same
size in \textbf{4}. The total frequency splitting is thus negative for
compounds \textbf{1}, \textbf{3} and \textbf{2} and positive for compound
\textbf{4}, where the spin-orbit coupling paramagnetic contributions almost
cancel each other out.

In the second and third set of compounds, molecules \textbf{5} to
\textbf{10}, which show large PV frequency splittings compared to the first
set, the relative ordering of the absolute values of the PV NMR frequency
splittings is
$\mathbf{5<6<7}$ and $\mathbf{8<9<10}$. If we regard them as two distinct
sets as shown in Figure \ref{fig:NMR}, the same is true for the PV energy
differences. For both properties, however, the three molecules containing
iodine display larger absolute values than the other three, and NWHFI shows
the largest absolute value with $\Delta\nu_{\mathrm{PV}} = -363~\upmu$Hz.
This comes also from the fact that the paramagnetic and spin-orbit coupling
contributions to the PV frequency splitting are of the same sign for
molecules of the second and third set, which enhances the increase in
frequency splittings.  All of the predicted $^{183}$W PV NMR frequency
splittings for members of the second and third series of molecules,
\textbf{5} to \textbf{10}, are significantly larger than the values
predicted for molecules \textbf{1} to \textbf{4}, and substituting any of
the halogens in the first series by hydrogen almost always leads to at
least an order of magnitude increase in the PV NMR frequency splittings.
The only exception to this is the substitution of iodine by hydrogen in
NWClFI, where there is an increase by a factor 4 ``only''. A possible
reason for this trend is the larger asymmetry (chirality) of the electronic
environment of the tungsten nucleus, introduced by substituting atoms of
different size and electronegativity.

It is also possible to analyze the impact of atomic substitution 
with respect to the different electronegativities ($\chi$) of the 
substituents, which is largest for fluorine and  smallest for 
hydrogen:
The relative sequence of electronegativities ($\chi$) of the 
substituents is $\chi \left( \mathrm{F} \right) > \chi \left( 
\mathrm{Cl} \right) > \chi \left( \mathrm{Br} \right) > \chi \left( 
\mathrm{I} \right) > \chi \left( \mathrm{H} \right) $ with $\chi 
\left( \mathrm{Cl} \right) \gtrsim \chi \left( \mathrm{N} \right) 
\gtrsim \chi \left( \mathrm{Br} \right)$. As mentioned earlier, this 
could explain, why in the first series of molecules there is a 
pronounced increase in the absolute values of both $\Delta 
E_{\mathrm{PV}}$ and $\Delta\nu_{\mathrm{PV}}$ upon fluorine 
substitution. It could also explain an increase in PV properties upon 
substitution of hydrogen for one of the three heavier halogens, but 
it is less clear, why there should be such a pronounced increase even 
when hydrogen is substituted for fluorine.

Regarding PV NMR frequency splittings, a comparison with $^{13}$C NMR
shielding tensors in CHBrClF and CHBrFI presented in Ref. \onlinecite{weijo:2005}
shows, that like the PV energy difference, the isotropic $^{183}$W NMR
shielding constants are of opposite sign in this series and the relative
ordering of the resulting frequency splittings is also changed: In
Ref. \onlinecite{weijo:2005} the $^{13}$C NMR shieldings in CHBrClF and CHBrFI are
similar in size with the ordering depending on the choice of density
functional. In the present study, the isotropic $^{183}$W PV NMR shielding
constant in NWBrFI is almost twice as large as that in NWBrClF. However,
for the values of $\Delta E_{\mathrm{PV}}$ reported in Refs.
\onlinecite{laerdahl:2000a,berger:2007}, the increase from CHBrClF to CHBrFI is
even more pronounced than the one reported herein for the $^{183}$W
shieldings in NWBrClF and NWBrFI.

\section{Conclusion}
We have studied PV effects for the NMR shielding tensor in chiral
tungsten compounds within the ZORA approach. We found that
PV NMR frequency splittings seem to be even more sensitive to
atomic substitution than PV energies, with an increase of
$\Delta\nu_{\mathrm{PV}}$ by three orders of magnitude from NWBrClI to
NWHFI! This sensitivity offers excellent prospects for the design of
thermodynamically stable compounds suited for a NMR experiment aiming at molecular PV, where it
would seem prudent to surround a heavy, NMR active nucleus in the chiral
center of a molecule with ligands providing a strongly heterogeneous
electronic environment.

\acknowledgments
Financial support by the Volkswagen Foundation and computer time provided
by the Center of Scientific Computing (CSC) Frankfurt are gratefully
acknowledged.  PS is indebted to the Alexander von Humboldt Foundation
(Bonn) for financial support during the stay at the Philipps University
of Marburg.

\end{document}